\documentclass[useAMS,usenatbib,usegraphicx]{mn2e}

\title[UV dust attenuation in spiral galaxies: the role of age-dependent 
extinction and of the IMF]{UV dust attenuation in spiral galaxies: the role of
age-dependent extinction and of the IMF}

\author[P. Panuzzo et al.]{P. Panuzzo,$^{1,2}$\thanks{E-mail: 
panuzzo@pd.astro.it} G. L. Granato,$^{1}$ V. Buat,$^{2}$ A. K. Inoue,$^{2,4}$
L. Silva,$^{3}$ \newauthor J. Iglesias-P\'aramo$^{2,5}$ and A. Bressan$^{1}$\\
$^{1}$INAF Padova, Vicolo dell'Osservatorio 5, I-35122 Padova, Italy \\
$^{2}$Observatoire Astronomique Marseille Provence,
Laboratoire d'Astrophysique de Marseille, 13012 Marseille, CEDEX 12, France\\
$^{3}$INAF Trieste, Via Tiepolo, Trieste, Italy\\
$^{4}$College of General Education, Osaka Sangyo University,
3-1-1, Nakagaito, Daito 574-8530, Japan\\
$^{5}$Istituto de Astrof{\'\i}sica de Andaluc{\'\i}a (CSIC), 18008 Granada,
Spain}

\pagerange{\pageref{firstpage}--\pageref{lastpage}} \pubyear{2006}

\begin{document}

\maketitle

\label{firstpage}

\begin{abstract}
We analyse the attenuation properties of a sample of UV selected
galaxies, with the use of the spectrophotometric model {\scshape Grasil}.
In particular, we focus on the relation between dust attenuation and the
reddening in the UV spectral region. We show that a realistic modelling of
geometrical distribution of dust and of the different population of stars can
explain the UV reddening of normal spiral galaxies also with a standard Milky
Way dust. Our results clearly underline that it is fundamental to take into
account that younger stars suffer a higher attenuation than older stars
(the {\it age-dependent extinction}) because stars are born in
more-than-average dusty environments.
In this work we also find that the concentration of young stars on the
galactic plane of spirals has a relevant impact on the expected UV colours,
impact that has not been explored before this paper.
Finally, we discuss the role of IMF in shaping the relation between
UV reddening and dust attenuation, and we show that a Kroupa IMF is more
consistent with observed data than the classical Salpeter IMF.
\end{abstract}



\section{Introduction}

One of the main quantities characterizing the physical and evolutionary status
of galaxies is their Star Formation Rate (SFR). Since long, many recipes to
estimate the SFR, using data taken in various spectral windows, either in the
continuum, or in emission lines, have been considered (e.g. 
\citealt{kenn98,hira03,panu03}). One obvious possibility is to use UV
radiation, since even in moderately actively star-forming systems, such as
normal spirals, this radiation is expected to be dominated by short lived
massive stars, which therefore should trace the past $10^8$ yr of star
formation activity. The drawback of this approach is that the UV radiation is
strongly affected by dust obscuration. The issue is particularly delicate
because stars are formed in more-than-average dense regions, and therefore they
are particularly affected by obscuration during their early evolution. Indeed
it is now well established that in IR bright galaxies, such as ULIRGs or sub-mm
selected high-z galaxies, the UV flux alone is hardly or not at all a reliable
indicator of the SFR (e.g. \citealt{gold02,flor99}), and even in normal spirals
it should be used with proper care \citep{bell02,bell01,buat96,buat02}.

In particular, when dealing with the dust obscuration in the UV domain, (i) it
has to be taken into account that the relative geometry of stars and dust
depends on the age of the stars ({\it age-dependent 
extinction}\footnote{Sometimes
called also age {\it selective} extinction.},
\citealt{silv98}), and (ii) it has to be understood if the dust optical
properties adopted to interpret the data are well representative of the
particular system. Actually, it is very difficult to disentangle from
observation of external galaxies the effects of dust optical properties from
those due to their complex geometry (and possibly those due to the star
formation history). It is therefore advisable to clearly distinguish the {\it
attenuation law} of a galaxy from the {\it extinction law} of the dust. While
the latter describes just the wavelength dependence of the optical properties
of the dust mixture, the former is the ratio (expressed in magnitudes) between
the observed and intrinsic
starlight, as a function of wavelength. This would be a direct measure of the
extinction law only if the stars were obscured by a thin screen of dust between
them and the observer, in general an extremely idealized situation.

For our own and a few nearby galaxies, the  extinction law of the dust can be
measured directly from observations of background stars, where indeed the dust
acts as a foreground screen. The differences found between the shapes of the
extinction curves of the Galaxy, the Large Magellanic Cloud and the Small
Magellanic Cloud below $\lambda\leq$2600 \AA\ \citep{fitz89} are often ascribed
to the different metallicities in these systems, covering the range $Z\sim 
0.1-1\ Z_\odot$. \citet*{calz94} have analysed the dust attenuation in
starburst galaxies. In this case, the derivation of the intrinsic extinction is
not direct, since one measures the integrated light of the whole system, where
stars and dust are mixed in a complex way. From the optical and UV spectra of a
sample of UV-bright starbursts they derive an average attenuation law ({\it
Calzetti law}) characterized by the absence of the 2175 \AA\ feature and a
far-UV slope (above the bump) shallower than that of the Milky Way extinction
law.

\citet*{gord97} argued that the observed shape of the Calzetti law can only be
explained with dust that lacks the 2175~\AA\ feature in its extinction curve.
However they only considered clumping of dust, not of stars, and assumed a
spatial distribution for stars independent of stellar age. Indeed a completely
different conclusion has been reached by \citet{gran00}, by means of the first
spectral synthesis model for the entire UV to IR domain taking into account the
effect of age-dependent extinction ({\scshape Grasil}, \citealt{silv98}). \citet{gran00}
(see also \citealt{panu03b}) have shown that this more realistic geometry
can explain the differences between the observed attenuation law of starburst
galaxies and the galactic extinction law, even adopting the galactic cirrus
optical properties. Note that this by no means demonstrates that the optical
properties of dust in starbursts are well represented by the average cirrus
dust in the Galaxy, but rather that any difference may be completely masked by
geometrical effects. In this interpretation, the attenuation law of a specific
stellar system arises, via the age-dependent extinction, from a complex 
blending of
optical properties of dust, star formation history and relative geometry of
stars and dust, including clumping of both components. The relative importance
of these ingredients is a function of the evolutionary status of the galaxy. In
very active systems the first ingredient is of little relevance, because much
of the intrinsic optical-UV starlight is produced by stars embedded in
molecular clouds, completely opaque at these wavelengths (see Section
\ref{sec_att_law}). Conversely, more
gently star forming galaxies, such as normal spirals, are expected to be better
candidates to investigate variations of dust properties, provided that a
sufficiently realistic geometry is used in their modelling.

Further clues on these complexities can be derived considering that different
classes of galaxies lie in different regions of the plane defined by the ratio
of IR and UV fluxes $F_{\rm IR}/F_{\rm UV}$ and the UV spectral slope $\beta$
(or, equivalently, a UV colour).

\citet*{meur99} found that UV bright starburst galaxies follow a well defined
correlation between the $F_{\rm IR}/F_{\rm UV}$ and $\beta$. These authors
noted that their relation implies an attenuation law with a slope around
1600 \AA\ very different from the Milky Way extinction law, and more similar
to the Calzetti law (that was derived using some galaxies belonging to the same
sample). Also \citet{char00} explained the $F_{\rm IR}/F_{\rm UV}$ vs $\beta$
relation in these objects using a featureless power-law extinction law.

If the correlation found by \citet{meur99} were a general property of galaxies,
it would allow to reliably estimate the SFR of galaxies from UV data alone.
However it is now established that this relationship is not a general property.
In particular, luminous IR galaxies have $F_{\rm IR}/F_{\rm UV}$ ratios orders
of magnitude higher than UV bright starbursts with the same $\beta$
\citep{gold02,burg05a}. Thus for these objects the procedure would lead to
severe underestimates of the actual SFR.

On the other hand, \citet{bell02} and \citet{kong04} collected data from
different UV space experiments and found that normal spirals are even redder
than the $F_{\rm FIR}/F_{\rm UV}$ vs $\beta$ relation found for UV bright
starbursts. \citet{kong04}, using the simple model by \citet{char00}, ascribed
this difference to the lower ratio between the recent over average star
formation activity. Note that they used the same featureless extinction law
assumed for starburst galaxies.

The {\it Galaxy Evolution Explorer} ({\it GALEX}) \citep{mart05} provides very
useful data to study this problem. In particular, the UV slope can be estimated
from data taken in the NUV and FUV {\it GALEX} bands. It is worth noticing that
the NUV filter, centered at 2310 \AA, is strongly affected by the presence of
the 2175 \AA\ bump in the MW extinction law. \citet{buat05} and \citet{seib05}
(see also \citealt{cort06}) confirmed with first {\it GALEX} observations that
normal spirals do not follow the $F_{\rm FIR}/F_{\rm UV}$ vs $\beta$ relation
found for UV bright starbursts and that the position of objects in the diagram
depend on the selection criteria.
Finally, \citet{calz05} found that also star-forming regions within NGC5194
do not follow the above starburst relation.

In this paper we use the relatively sophisticated spectral synthesis model
{\scshape Grasil} to interpret the UV and FIR properties of normal spiral
galaxies. We
will show that these properties can be explained without invoking strongly
non-standard dust, provided that a sufficiently detailed (but realistic)
description of the geometry of the system is employed. Conversely, these
properties would require in some sense extreme assumptions about the dust
composition if a simplistic geometry were adopted. In particular, we will show
the effect of taking into account that in normal spirals stars younger than
$\sim 10^8$ yr are confined in a thin disc with a smaller scale-height than
older stars and diffuse medium. The different attenuations suffered by young
and old stars have a relevant impact on UV colours, a fact
that has not been considered in any of the above mentioned works (see
\citealt{buat96} for an analysis of the effect on dust attenuation estimation).

We will also discuss the role of the initial mass function (IMF) showing that
a IMF with less massive stars than a Salpeter one can be more consistent with
data.
In particular, we will show that this makes the cirrus more efficient in
attenuating young stars and reddening the final SEDs.

This paper has the following structure.
In Sections \ref{sec_data} and \ref{sec_obsirx} we describe the 
data sample we consider for our analysis. In Section \ref{sec_grasil}
we describe our spectro-photometric model for dusty galaxies, {\scshape Grasil}.
In Section \ref{sec_geo} we apply our code to study different
geometrical configurations of dust and stars and the relevance
of an age-dependent extinction to explain the data. Furthermore,
in section \ref{sec_att_law} we discuss the attenuation law
resulting from our models, and in section \ref{sec_imf} we discuss the role
of the IMF. Finally, in Section \ref{sec_discus} we summarise and discuss the
results of this work.


\section{The \textit{GALEX} NUV selected sample}

\subsection{The data}
\label{sec_data}

The galaxy sample we consider in our analysis is derived from the first
observations
of the {\it Galaxy Evolution Explorer} mission and from IRAS observations. The
{\it GALEX} data are photometric measurement in two ultraviolet bands: the
Far-UV band (FUV, 1350--1750 \AA, $\lambda_{\rm mean} =1520$ \AA) and the
Near-UV band (NUV, 1750--2750 \AA, $\lambda_{\rm mean} =2310$ \AA).

The sample was presented in \citet{buat05} and \citet{igle06}; it consists in
59 galaxies selected on the basis of the NUV flux ($m_{\rm NUV} \leq 16$ mag, 
AB system). Here, we only summarize the main information about the 
sample, more details on the sample can be found in \citet{igle06}.

The galaxy fluxes in the sample were extracted from the {\it GALEX} All-Sky
Imaging Survey \citep{mart05} and corrected for the foreground Galactic
extinction using the \citet{schl98} dust map and the \citet*{card89} extinction
curve. Sources were identified using the SIMBAD, 2MASS\footnote{The Two Micron 
All Sky Survey is a joint project of the University of Massachusetts and the 
Infrared Processing and Analysis Center/California Institute of Technology, 
funded by the National Aeronautics and Space Administration and the National 
Science Foundation.} and PGC catalogs; the
same catalogs were used to determine the morphological types of the sample
galaxies. Two ellipticals, four Seyferts and one QSO were discarded. The FIR
fluxes of the remaining sources were taken from the IRAS Faint Source Catalog
(version 2) \citep{mosh90} and the Scan Processing and Integration Facility
(SCANPI) at 60 and 100 $\umu$m. We excluded from the sample multiple UV sources
not resolved as single objects by IRAS and sources contaminated by cirrus.
Complementary data in visible are taken from HyperLeda\footnote{HyperLeda
database is available at the URL {\tt http://leda.univ-lyon1.fr}}, while $H$
magnitudes from 2MASS.


\subsection{The observed IRX-$\bmath{\beta}$ diagram}
\label{sec_obsirx}

In this paper we will focus on the relation between
$F_{\rm IR}/F_{\rm UV}$ and $\beta$, already called IRX-$\beta$ diagram. For
this purpose we will express the ratio $F_{\rm IR}/F_{\rm UV}$ with the ratio
between $F_{\rm TIR}$, i.e. the total dust emitted flux estimated from IRAS 60
and 100 $\umu$m fluxes following the prescription of \citet{dale01}, and
$F_{\rm FUV}$, the observed flux in the {\it GALEX} FUV band. Moreover, we will
express the UV spectral slope $\beta$ with the NUV$-$FUV colour.

The IRX-$\beta$ diagram as seen by {\it GALEX} for these NUV selected galaxies
was presented in \citet{buat05} and \citet{burg05b}. Here we would like to
underline the effect of the star formation activity on the diagram.

As already mentioned, \citet{kong04} suggested that the displacement between
normal spiral galaxies and starburst in the $F_{\rm IR}/F_{\rm UV}$ vs UV
colour is due to the different ratio between the recent over average star
formation activity. This ratio is quantified by the birthrate parameter $b$
\citep{kenn94}, defined as the ratio between the  present star formation rate
(SFR) and the mean SFR ($\langle{\rm SFR}\rangle$) in the past:
\begin{equation}
b\equiv \frac{\rm SFR}{\langle{\rm SFR}\rangle} ~.  
\end{equation}

As for the present star formation rate, it can be derived from the NUV
luminosity, properly corrected for dust attenuation using the ratio IR/UV. From
our SSP library (see section \ref{sec_grasil}), we computed the NUV luminosity
of a constant SFR lasting 100 Myr, assuming a Salpeter IMF from 0.1 to 100
M$_\odot$. The relation between SFR and $L_{\rm NUV}$ we got is:
\begin{equation}
\log \frac{L_{\rm NUV}}{1 \rm L_\odot}= \log \frac{\rm SFR}{1 \rm M_\odot/{\rm yr}}+9.356~.
\label{eq_nuvcalib}
\end{equation}

As discussed by many authors (e.g. \citealt{buat96,buat99,panu03}), the ratio
IR/UV is a good estimator of the UV attenuation. We used here the relation
proposed by \citet{buat05}:
\begin{equation}
A_{\rm NUV}=-0.0495x^3+0.4718x^2+0.8998x+0.2269
\end{equation}
where $x=\log (F_{\rm TIR}/F_{\rm NUV})$. Finally, the mean star formation rate
can be derived from the $H$ luminosity as shown by \citet{bose01}.

The resulting mean value of $b$ for our NUV selected sample is $b\sim 0.5$, 
typical of normal late type spirals, confirmed by
the morphological classification (see \citealt{igle06} for discussion). Also
the infrared colours of our sample objects indicate a mild level star formation
activity.

\begin{figure}
\includegraphics[angle=270,width=\hsize]{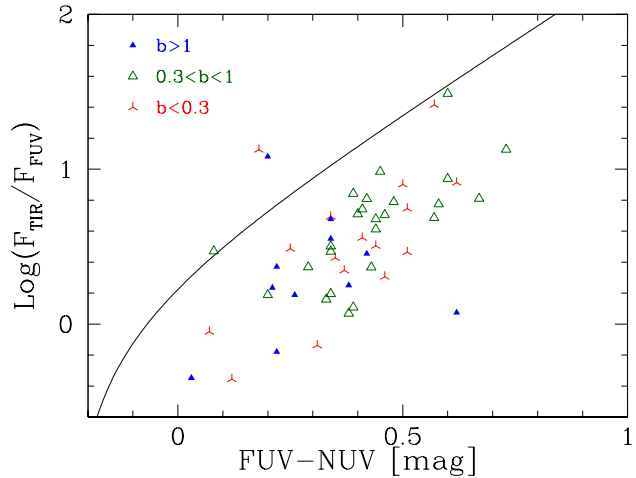}
\caption{IRX-$\beta$ diagram for the NUV selected sample. Filled triangles are 
objects with a birthrate parameter $b$ larger than 1.0; empty triangles
represent objects with $0.3 < b < 1.0$, while stars represent objects with
$b < 0.3$. The solid line represents the Meurer et al's UV bright 
starburst sequence converted into GALEX bands (see the text for more datails).}
\label{fig_irxuv_oss}
\end{figure}

The position of sample galaxies in the IRX-$\beta$ diagram is shown in figure
\ref{fig_irxuv_oss}, where we have used different symbols for objects having
$b<0.3$, $0.3 < b < 1.0$ and $b > 1$. 

In the same diagram we show the UV bright starburst sequence of the 
\citet{meur99} sample where the UV spectral index was converted into the
FUV$-$NUV colour following \citet{kong04}. Thus the starburst relation is given
by:
\begin{equation}
\frac{F_{\rm TIR}}{F_{\rm FUV}}=10^{0.4+1.87({\rm FUV-NUV})}-0.95~.
\label{equ_meurer_rel}
\end{equation}

It is worth noting that (i) most of the galaxies have a redder FUV$-$NUV colour
than what expected by the starburst relation, regardless their birthrate
parameter value; (ii) galaxies with a small birthrate parameter are not
systematically redder than more active galaxies. These two statements
contrast with what predicted by \citet{kong04}, i.e. that galaxies with
birtrate $b \sim 1$ should follow the starburst relation, while galaxies with
$b \geq 0.3$ should be redder. There are some explanations for this
disagreement. As discussed in \citet{burg05b}, the NUV flux used in the
computation of the birthrate parameter is produced by all stellar populations
with an age smaller than $\sim$ 100 Myr, so the present SFR is estimated with
an average over the last $\sim$ 100 Myr. In presence of a star formation
history with significant variations in this interval of time, the SFR estimated
with the NUV flux could be not representative of the instantaneous SFR.
However, the observed FUV$-$NUV colour is due to a complex interplay between
the star formation history {\it and} the effect of the dust attenuation (as
discussed in the following sections), so it is not clear if our birthrate
parameter estimation is fully representative for the IRX-$\beta$ diagram.
Indeed, \citet{cort06} analized the relation between the distance from the
starburst relation and the birthrate parameter computed using the H$\alpha$
flux (thus using an estimate of instantaneous SFR), but they found only a weak
correlation for normal star-forming galaxies.

Another possible factor in this disagreement is that the equation
\ref{equ_meurer_rel} for the Meurer et al. sample was obtained by translating
the Meurer's definiton of $\beta$ into the FUV$-$NUV colour, and not by
observing the same sample with {\it GALEX}. Indeed, \citet{calz05} proposed an 
alternative conversion of the 
UV spectral index into the FUV-NUV which provides a redder colour
for the UV bright starburst sequence.
Thus, we think that the equation \ref{equ_meurer_rel} should be verified
with {\it GALEX} observations.


\section{Spectral energy distribution of galaxies: G{\sevensize\bf RASIL}}
\label{sec_grasil}

We compute the SEDs of galaxies using our code {\scshape Grasil}. {\scshape 
Grasil}\footnote{The code
is available at the URL {\tt http://web.oapd.inaf.it/granato} or {\tt
http://adlibitum.oat.ts.astro.it/silva/default.html}. It can be run also with
an user friendly web interface, {\scshape Galsynth}, available at the URL {\tt 
http://web.oapd.inaf.it/galsynth}.} is a population synthesis model to predict
the time dependent SEDs of galaxies from far-UV to radio, including
state-of-the-art treatment of all relevant aspects dust reprocessing
\citep{silv98,gran00,vega05}, production of radio photons by thermal and
non-thermal processes \citep*{bres02} and nebular lines emission
\citep{panu03}. We refer the reader to the original papers for details, but
we summarize for convenience the main features.

The starting input for {\scshape Grasil} is the history of star formation and
chemical
enrichment histories of the system. This is computed by an external code, which
can result from a complex scenario for the formation of galaxies in a
cosmological context (e.g. \citealt{gran00,gran01,gran04}), or a standard
chemical evolution model for the formation of a single galaxy. Here, as in
\citet{silv98}, we follow the latter approach. Our chemical evolution model is
a standard implementation of a one zone open model including infall of
primordial gas (see \citealt{silv98} and references therein). The parameters
regulating the star formation history are the baryonic mass of the galaxy
($M_\mathrm{G}$), the gas infall time scale ($\tau_\mathrm{inf}$), and the star
formation efficiency ($\nu_\mathrm{sch}$) of the assumed Schmidt law.

Once the star formation and chemical enrichment histories of a galaxy are
given, {\scshape Grasil} computes the interaction between the stellar radiation and dust
using a relatively realistic and flexible geometry for both stars and dust.

One of the most important distinctive features of {\scshape Grasil} is that it
included,
for the first time, the effect of {\it age-dependent extinction} of stellar
populations (younger stellar generations are more affected by dust
obscuration), due to the fact that stars form in a denser than average
environment. In particular, new stars are born inside MCs and progressively get
rid of them, either because they escape or because the clouds are destroyed.
This is described in {\scshape Grasil} as follows. If $t_{\rm esc}$ is the
timescale for the process, the fraction of starlight radiated inside the
clouds at time $t$ after they formed is given by
\begin{equation}
F(t)=\left\{\begin{array}{lll} 1&~~&\hbox{for } t<t_{\rm esc}\\
2-t/t_{\rm esc}& &\hbox{for } t_{\rm esc}<t<2t_{\rm esc}\\
0& &\hbox{for }t>t_{\rm esc}
\end{array}\right.
\end{equation}
In practise, 100\% of the stars younger than $t_{\rm esc}$ are considered to
radiate inside the MCs, and this percentage goes linearly to 0\% in $2 
t_{\rm esc}$. The timescale $t_{\rm esc}$ is a fundamental parameter, which was
found longer in an handful of well studied local starburst than in normal
disc-like galaxies by a SED fitting procedure \citep{silv98,silv99}.

The galaxies' geometry in {\scshape Grasil} is generally described as a
superposition
of an exponential disc component and a bulge component, the latter modelled
by a King profile.

The stars and dust density in the exponential disc component is give by the
following equation:
\begin{equation}
\rho=\rho_0\cdot \exp(-R/R_{\rm d}) \cdot \exp(-|z|/z_{\rm d})
\label{eq_expdisc}
\end{equation}
with $\rho_0$ the central density, $R_{\rm d}$ and $z_{\rm d}$ the scalelength
along the radial and normal directions. The code allows to specify different
values for $R_{\rm d}$ and $z_{\rm d}$ for stars and dust.

Up to now, we have only considered the contribution of molecular
clouds to the age-dependent extinction, which is indeed the most
prominent, at least in very active systems. In this paper we show
that to interpret the UV spectral shape of spiral galaxies, it is
necessary a further step, namely to consider the fact that MCs are
concentrated in the galactic plane, and therefore the same is true
for young stars already free from their parent MC. As a result,
though free from MCs obscuration, these stars are more extincted
by the diffuse medium than older populations.

We model this geometrical configuration by dividing the star content of the
galaxy into two
populations, according to their age. If $t_{\rm thin}$ is the timescale on
which young stars stay concentrated on the galactic plane, we assume two
different scalelengths sets ($R_{\rm d}$ and $z_{\rm d}$) for stars younger or 
older than $t_{\rm thin}$.

The total gas mass (diffuse+MCs) of the galaxy at time $T_{\rm G}$ is given by
the chemical evolution model. The relative fraction of molecular gas is a free
parameter of the code, $f_{\rm mc}$. The total molecular mass, $M_{\rm mc}$ is
then subdivided into spherical clouds of mass and radius, $m_{\rm c}$ and
$r_{\rm c}$. Then, the radiative transfer of starlight through the MCs and
diffuse ISM is solved.

The ratio between gas mass and dust mass $G/D$ is assumed to scale linearly
with the metallicity of the residual gas and $G/D=1/110$ for $Z=Z_\odot$. This
ratio plus $m_{\rm c}/r_{\rm c}^2$ determine the optical depth of the clouds.
Note that in {\scshape Grasil}, the predicted SED depends on $m_{\rm c}$ and $r_{\rm c}$
only through the combination $m_{\rm c}/r_{\rm c}^2$, which is the true free
parameter\footnote{In \citet{silv98} as well as in the present paper we found
sufficient to consider
a single {\it population} of MCs all with the same parameters, but {\scshape Grasil}
treats in general many populations each with different properties. Of course,
this would add more parameters.}.

For the dust composition, we adopt a mixture of graphite and silicate
grains and PAHs. In general, the size distributions are adjusted to match the
extinction and emissivity properties of the local diffuse ISM in the Galaxy
(for more details see \citealt{silv98} and \citealt{vega05} for the improved
model of PAHs), but we have also considered other possibilities (see section
\ref{sec_geo}).

The SSPs included in {\scshape Grasil} are based on the Padova stellar models,
and cover a
large range of ages and metallicities. Starlight reprocessing from dust in the
envelopes of AGB stars is included directly into the SSPs, as described by
\citet*{bres98}.


\section{Interpretation of \textit{GALEX} IRX-$\bmath{\beta}$ diagram}

To interpret the IRX-$\beta$ diagram, we computed different sets of disc
galaxy models.

\begin{figure}
\includegraphics[angle=270,width=\hsize]{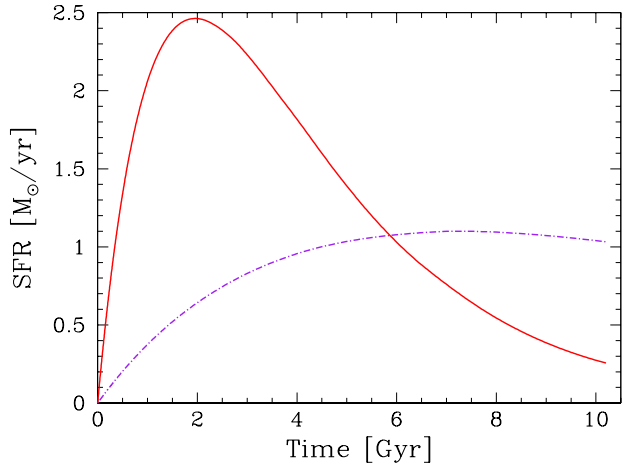}
\caption{Star formation histories used in our models. Solid line is for a
birthrate parameter $b=0.25$, dot-dashed line is for $b=1.25$.}
\label{fig_sfr}
\end{figure}

For simplicity, here we consider only two star formation histories (Figure 
\ref{fig_sfr}) with birthrate parameters $b=1.25$ and $b=0.25$. These values
span most of the observed spread of the birthrate parameters in the sample. We
computed the SED at a galactic age of 10 Gyr for all models.

\subsection{The role of dust and stars geometry}
\label{sec_geo}

To illustrate the effects of relative star-dust geometry, and in particular the
age-dependent extinction, we consider four sets of model. For all models of
this section (except the last one), we assume a Salpeter initial mass function
(IMF).

\begin{figure}
\includegraphics[angle=270,width=\hsize]{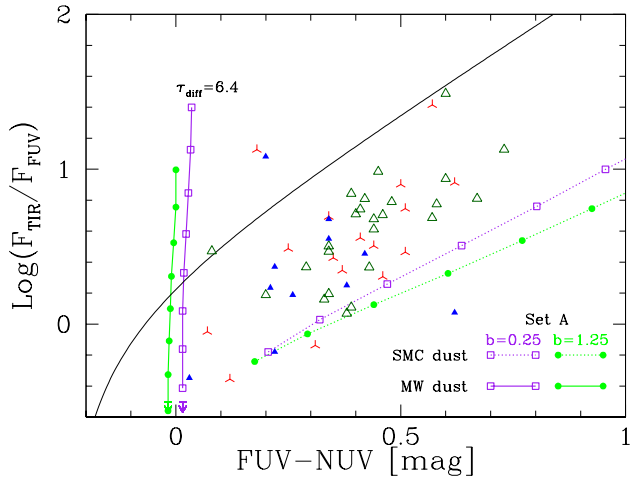}
\caption{Comparison between the NUV selected sample galaxies (as in fig
\ref{fig_irxuv_oss}) and set A models. Connected dots are for models with
increasing optical depth of the diffuse medium.}
\label{fig_mod_disc}
\end{figure}

In the first set (set A) we neglect any age-dependent extinction, i.e. all the
dust is assumed to stay in the diffuse medium (cirrus), without molecular
clouds, and the stars are smoothly distributed within this medium independently
of their age. The geometrical distributions of stars and dust are exponential
discs (for clarity, we do not include a bulge component), described by eq.
\ref{eq_expdisc} with scale values reported in table
\ref{tab_scaleval}. The set consists of a sequence of models with increasing
gas (and therefore dust) content\footnote{{\scshape Grasil} allows to specify
at wish the
gas content of the galaxy, or to use the value predicted by the chemical
evolution code; in set A we specify the gas content by hand in {\scshape
Grasil}.}. The
parameter describing this sequence is the cirrus optical depth at 1$\umu$m from
the center along the polar axis, ranging from 0.05 to 6.4, and doubling between
one model and the other. The models were computed for two choices of dust
composition; one reproducing the MW extinction law, the other reproducing the
extinction law of SMC (dust composition from \citealt{wein01}). The position of
models in the IRX-$\beta$ diagram is shown in figure \ref{fig_mod_disc} (models
connected with solid lines have a MW dust; those connected with dotted lines
have a SMC dust), where NUV and FUV magnitudes have been computed taking into
account the filter transmissions. All models in figure \ref{fig_mod_disc} are
face-on, i.e. with the polar axis parallel to the line of sight. The two arrows
shows the FUV$-$NUV colour of dust-free models for the two star formation
histories. Note that the dust-free FUV$-$NUV colour is redder than
the zero-extinction color for the UV bright starburst sequence due to the assumed
continuous star formation.

\begin{table}
\caption{Geometrical parameters for the models. Note that set B models do not
have diffuse dust, thus they do not depend on $R_{\rm d}$ or $z_{\rm d}$.}
\label{tab_scaleval}
\begin{tabular}{lcc}
\hline
  &Set A & Set C and D\\
\hline
$R_{\rm d}^{\rm stars}$ & 1 Kpc   &1 Kpc (young), 2 Kpc (old)  \\
$z_{\rm d}^{\rm stars}$ & 50 pc   &50 pc (young), 400 pc (old)  \\
$R_{\rm d}^{\rm dust}$  & 1.5 Kpc &1.5 Kpc \\
$z_{\rm d}^{\rm dust}$  & 50 pc   &100 pc  \\
\hline
\end{tabular}
\end{table}

Models with MW dust move in the diagram at nearly constant FUV$-$NUV colour as
the dust content increases, while those with SMC dust become redder and redder.
The lack of reddening for the MW dust case is due to the presence of the 2175
\AA\ bump in the extinction law that affects strongly the NUV band. In fact,
the extinction in NUV is actually higher than in the FUV
($A_{\rm NUV}/A_{\rm FUV}=1.14$) for a MW dust. From the figure, it is clear
that a dust mixture somewhat intermediate between the MW and the SMC and more
similar to the latter could reproduce the data points. A similar result was
found by \citet{witt00}, investigating the variation of the UV spectral index.
They also found that this result does not depend on the relative geometry of
dust and stars, which however was always age independent.

\begin{figure}
\includegraphics[angle=270,width=\hsize]{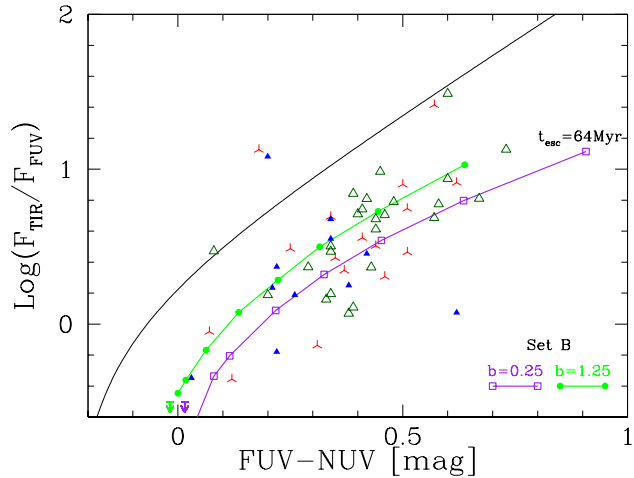}
\caption{Comparison between NUV selected sample galaxies (as in fig
\ref{fig_irxuv_oss}) and models with varing $t_{\rm esc}$ (set B). Connected
dots are for models with $t_{\rm esc}=$ 0.5, 1, 2, 4, 8, 16, 32 and 64 Myr (the
reddest models).}
\label{fig_irxuv_mc}
\end{figure}

In the second set of models (set B) we begin to explore the role of the
age-dependent extinction due to molecular clouds. In this set we try to reproduce
the observed IRX-$\beta$ diagram of the UV selected sample, assuming that all
the gas is arranged in clouds obscuring young stars with no gas left in the
cirrus. Here the sequence consists in models with increasing escape time, while
the total amount of gas is that predicted by the chemical evolution code. The
results are shown in figure \ref{fig_irxuv_mc}: solid lines connect models
(represented by circles or empty squares) with the same star formation history
and gas amounts; the escape time increases from 0.5 to 64 Myr doubling between
one model and the other. The optical depth of MCs in the models is $\tau_{\rm
MC}=20$ at 1 $\umu$m; however, the position on the IRX-$\beta$ diagram does not
depend on the value of $\tau_{\rm MC}$ provided that $\tau_{\rm MC}\ga 1.5$,
because MCs are completely opaque at the {\it GALEX} wavelengths. Above this
value, changes in the optical depth of MCs affect only the shape of the IR SED.
It is also important to notice that the position of this set on the IRX-$\beta$
diagram are independent of the dust properties. In fact, FUV and NUV fluxes
depend only on the escape time and the star formation history, while the
$F_{\rm TIR}$ is nothing more than the bolometric luminosity of stars inside
MCs.

To reproduce the reddest objects, very large values of the timescale with which
new stars get rid of their parent MCs are required, ($t_{\rm esc} \geq 30$
Myr). This solution is unsatisfactory on astrophysical grounds, since stars
whose lifetime is about one order of magnitude shorter than this are commonly
observed in optical bands in quiescently star forming galaxies. Moreover, the
IR dust emission for these models is hotter than the observed SED: for all
these models $\log (F_{60\umu \rm m}/F_{100\umu \rm m}) > -0.3$, while the
observed mean value is $\log (F_{60\umu \rm m}/F_{100\umu \rm m}) \simeq -0.45$
with objects having down to $\log (F_{60\umu \rm m}/F_{100\umu \rm m}) \simeq 
-0.8$. This is a consequence of the assumed geometrical configuration of the
dust that feels a stronger radiation field than in the diffuse medium,
resulting in a higher temperature and hotter IR emission.

Our interpretation of the above results is that the assumption of
age-independent extinction for all star populations outside MCs is not
satisfactory. The very long escape timescale from parent clouds tries to mimic
the fact that in real spiral galaxies stars younger than $\sim 10^8$ yr are
still very much concentrated in the galactic plane, and therefore they are more
affected by extinction from the diffuse ISM than older ones.

Thus, in the third set of models (set C) we explored the effect of the smaller
scale-height of young star distribution in the disc. Here we assume that the
dust is present only in the cirrus (no MC, like set A) and that stars younger
than an age $t_{\rm thin}$ have a smaller scale-height than older stars, while
the dust scale-height is in between. The scale-heights of different components
are reported in table \ref{tab_scaleval}. We computed SED models for different
values of $t_{\rm thin}$, ranging from 25 Myr to 200 Myr. As in set A, we
computed sequences of models with increasing dust content; the cirrus optical
depth at 1$\umu$m from the center along the polar axis goes from 0.05 to 6.4,
and doubling between from one model and the other. The models were computed for
the same dust compositions of set A.

\begin{figure}
\includegraphics[angle=270,width=\hsize]{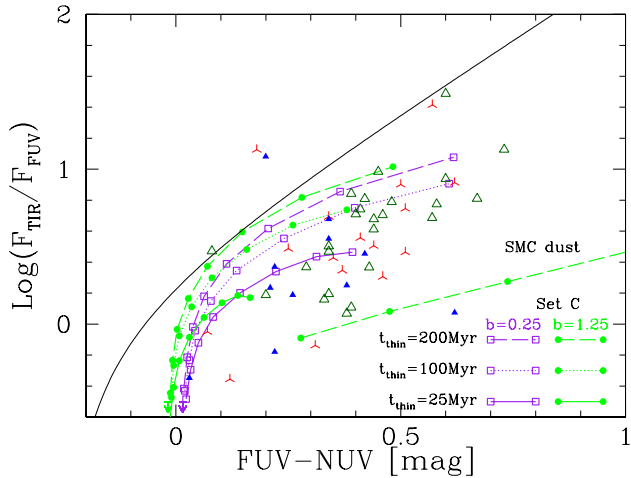}
\caption{Comparison between NUV selected sample galaxies and models with 
age-dependent scaleheight (set C). Connected dots are for models with increasing
optical depth of the diffuse medium.}
\label{fig_irxuv_sal}
\end{figure}

The comparison between the NUV selected sample and set C models is shown in
figure \ref{fig_irxuv_sal}. For clarity, we showed the MW dust models and only
one SMC dust model sequence. It is worth noting that, contrary to what happens
in the set A models, increasing the diffuse medium optical thickness 
produces a reddening also in MW dust models. The results show that when the 
optical depth is low the UV SED is dominated by the young population, and the
models move at constant FUV$-$NUV as the optical depth increases because the
geometry is
mixed. When the optical depth is high enough to attenuate considerably
the young population, the contribution of stars older than $t_{\rm thin}$
(that have a larger scale-height) becomes not negligible and the total
SED becomes redder. However, models are still bluer than the observed sample.

It should also be noted that models with a larger value of $t_{\rm thin}$
have a larger $F_{\rm TIR}/F_{\rm NUV}$ (for the same value of FUV$-$NUV).
In fact, if $t_{\rm thin}$ is larger, the young population enbedded near the
galactic plane is more luminous, thus a larger attenuation is needed to
make the old stellar population contribution non negligiable.

In the last set of models (set D) we consider the combined effect of the
smaller scale-height of young star distribution in the disc and the
age-dependent extinction due to molecular clouds.
Here we assume the same dust-star configuration used in set C, i.e.
with young stars with a small scale-height, but with the presence of 
molecular clouds with an escape time fixed to $t_{\rm esc}$=4 Myr.
Again we computed sequences of models with increasing dust content in the cirrus
(dust content of MCs is fixed). The cirrus optical depth at 1$\umu$m from 
the center along the polar axis goes from 0.05 to 6.4 
(except for the first models where we assume no cirrus), and 
doubling between from one model and the other.

\begin{figure}
\includegraphics[angle=270,width=\hsize]{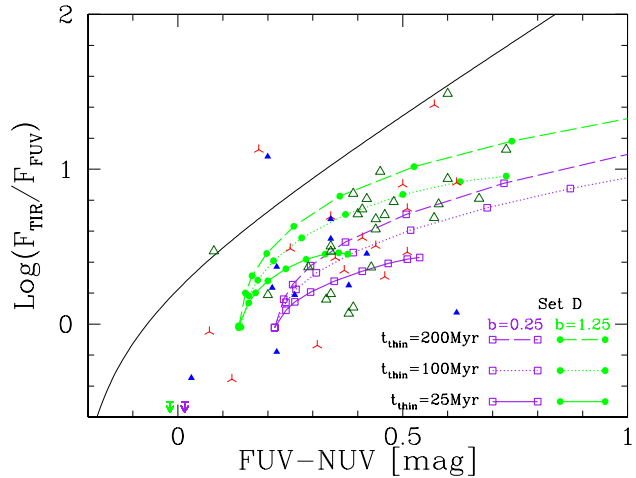}
\caption{Comparison between NUV selected sample galaxies and models with
age-dependent scale-heights and molecular clouds with $t_{\rm esc}$=4 Myr (Set
D). Connected dots are for models with increasing optical depth of the diffuse
medium.}
\label{fig_irxuv_sal_mc}
\end{figure}

It is worth noting that the presence of MCs increases the effect of cirrus dust
and the models are more compatible with observations. In fact, the MC
obscuration removes a significant fraction of UV luminosity from the young
population; thus the optical thickness of cirrus needed to attenuate the young
population outside MCs enough to make relevant the contribution of older stars
is smaller than in set C, making the models redder.

In conclusion, in this section we have shown that inferences on the
dust optical properties of galaxies derived from the analysis of
the IRX-$\beta$ diagram are dangerous, if the models are too schematic.
In particular, the common (and unrealistic) assumption that
starlight suffer the same reprocessing independently of the star
age is strongly misleading.


\subsection{The attenuation law}
\label{sec_att_law}

\begin{figure}
\includegraphics[angle=270,width=\hsize]{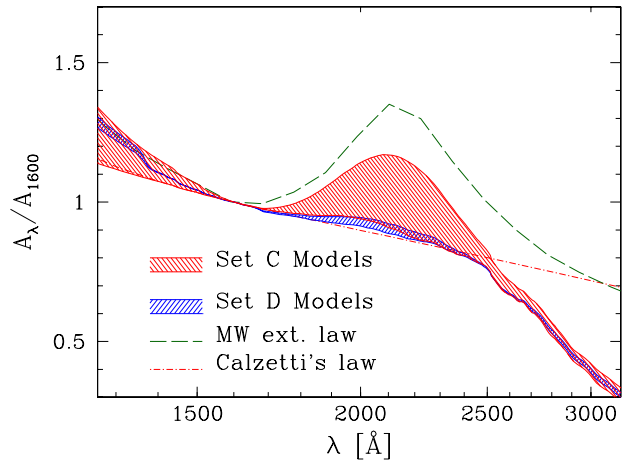}
\caption{The attenuation curves for simulated galaxies of sets C and D with the
attenuation law from \citet{calz94} (dot-dashed line) and the MW
extinction law (long-dashed line). All curves are normalized to unity at
$\lambda=1600$ \AA . Our data enable us to study the attenuation curves only in the UV.
Thus we cannot confirm nor disprove the deviations predicted by our models
with respect to the MW extinction law and the Calzetti law above 2500 A.}
\label{fig_curv_att}
\end{figure}

For a better understanding of the issue, we also analysed the attenuation law
resulting from our models. The attenuation law $A_\lambda$ in simulated
galaxies is thus defined as the difference in magnitudes of the luminosity
$L_\lambda$ of models with and without dust.

Figure \ref{fig_curv_att} compares the attenuation curves for
simulated galaxies of sets C and with the attenuation law from \citet{calz94}
and the MW extinction law. All curves are normalized to unity at 
$\lambda=1600$ \AA. The shadowed region shows the range of variation of the
attenuation laws of the models with the variation of the parameters; the
attenuation laws plotted in fig. \ref{fig_curv_att} are only for models with MW
dust.

For set C models, the one with the larger bump in the attenuation law
is the model with the smaller dust content. As the optical depth of dust
increases, the bump strength decreases until it vanishes completely for highest
optical depths. In fact, as the optical depth of dust increases, a growing
fraction of the observed UV is produced by the older stellar population; this
emission can fill the absorption feature in the SED of young population.

For set D models, indeed, the bump in attenuation law vanishes. In fact, in
this case the attenuation is mostly due to MCs that are opaque at these
wavelengths. Consider a system with two stellar populations, younger
stars inside MCs and stars outside. The observed luminosity $L_\lambda$ is
given by:
\begin{equation}
L_\lambda = (f^{\rm y}_\lambda 10^{-0.4\cdot A^{\rm y}_\lambda} + 
(1-f^{\rm y}_\lambda) 10^{-0.4\cdot A^{\rm o}_\lambda})\cdot 
L^{\rm em}_\lambda \, ,
\end{equation}
where $f^{\rm y}_\lambda$ is the luminosity fraction of stars inside MCs,
$A^{\rm y}_\lambda$ is the attenuation of stars inside MCs, 
$A^{\rm o}_\lambda$ is the one for stars outside, and $L^{\rm em}_\lambda$ is
the unextincted luminosity
If the optical depth of MCs is very high and that toward
the diffuse star population is very low, we have 
$L_\lambda  \simeq (1-f^{\rm y}_\lambda) L^{\rm em}_\lambda$. Thus
$A_\lambda$ is mostly independent of
the dust properties and depends mostly on the luminosity fraction.

The prediction of a decreasing bump strength with increasing attenuation
was also found in other works in literature (e.g. \citealt{gran00}), even in
the case where an uniform stellar population is assumed 
(\citealt{witt00,pier04}).
Note also that \citet{burg05b} found that the SEDs 
of NUV selected normal spiral galaxies show
an attenuation law with a reduced 2175 \AA\ bump. However, the above
authors found that the bump strength seems to increase with increasing
attenuation. This is probably due to the presence in their
sample of a number of IR selected galaxies with a blue FUV$-$NUV colour
and high dust attenuation.

Finally, it could be noted that the attenuation laws
from models decreases more steeply than the MW extinction law  
beyond 2500 \AA. This is due to the contribution (increasing with the wavelength) from 
older stars which suffer a much smaller attenuation than young stars
thus decreasing the attenuation of the whole system.
At this stage we cannot verify whether such deviation is actually
observed in our galaxies, as a comparison with data would require
assumptions on the intrinsic, dust--free SEDs of normal spirals, which,
at optical wavebands, strongly depend on their star formation history.


\subsection{The role of IMF}
\label{sec_imf}

We also explored the possibility of using a different initial mass function to
interpret the observations.

Several works (see \citealt*{port04} and references therein) claim that the
properties of normal disc galaxies are not well reproduced using a Salpeter
IMF. In particular, with a Salpeter IMF, chemical evolution models for
disc galaxies produce higer mass to light ratios than the observed ones, and a
too
high efficiency in metal production. These problems can be alleviated by a IMF
having a shallower slope at low masses and a steeper one at high masses. Among
others, \citet{krou98} proposed an IMF with the above characteristic, defined
as (we use the formulation given by \citealt{port04}):

\begin{equation}
\frac{dn(m)}{d\log m} \propto \left\{
\begin{array}{lcl}
m^{-0.5} &~~~~~ &\hbox{for } 0.1 < m < 0.5 ~{\rm M_\odot}\\
m^{-1.2} & &\hbox{for } 0.5 < m < 1 ~{\rm M_\odot}\\
m^{-1.7} & &\hbox{for } 1 < m < 100 ~{\rm M_\odot}
\end{array}
\right.
\end{equation}

We computed an SSP spectra library using the above formulated Kroupa IMF and
compared the resulting UV SED evolution with the one obtained using a 
Sapleter IMF.
In the {\it GALEX} spectral window, the UV SED of young SSPs is dominated
by the more massive stars ($m \geq 30$M$_\odot$), whose {\it GALEX}
colour does not depend on the mass. As a consequence, the UV 
spectral slope (or UV colours) of young SSPs with Salpeter and 
Kroupa IMF are very similar, despite the different high mass slope. 
With our SSP spectra we found that the FUV$-$NUV colour in the two IMF cases
always differs less than 0.03 mag, for all stellar ages younger than $10^8$ yr.

\begin{figure}
\includegraphics[angle=270,width=\hsize]{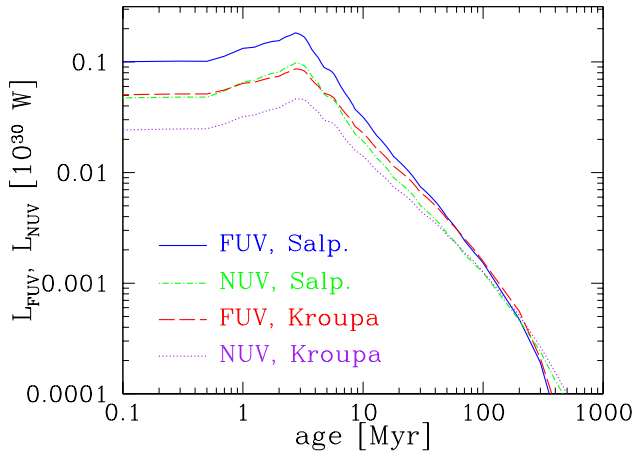}
\caption{Time evolution of NUV and FUV luminosities of a simple stellar 
population
of solar metallicity, with Salpeter and Kroupa IMF. The initial total mass of
the population is normalized to 1 M$_\odot$.}
\label{fig_fluxssp}
\end{figure}

However, the difference in the IMF slope at high masses gives
a different evolution of the UV luminosity between the two IMFs.
Indeed, figure \ref{fig_fluxssp} shows that the UV flux for a
Kroupa IMF decreases with the age of the population with a slower
rate than that corresponding to a Salpeter IMF, as $t^{-1.13}$ and
$t^{-1.43}$ respectively, between 3 and 100 Myr. As a consequence,
the FUV$-$NUV dust-free colour of a galaxy with $b=0.25$ is around 0
assuming a Salpeter IMF (figure \ref{fig_irxuv_sal}), and 0.2 with
a Kroupa IMF.

\begin{figure}
\includegraphics[angle=270,width=\hsize]{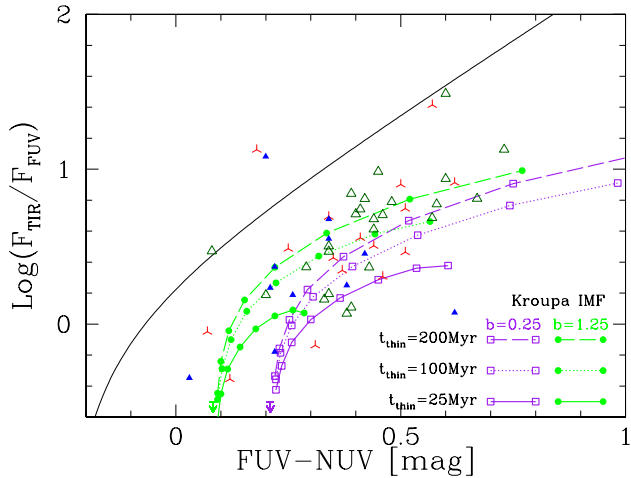}
\caption{Comparison between NUV selected sample galaxies and models 
with age-dependent scaleheight and Kroupa IMF.}
\label{fig_irxuv_kro}
\end{figure}

Thus we re-computed the spiral galaxies models with the age-dependent
scale-height (like set C in the previous section) but adopting a Kroupa 
IMF\footnote{We neglect here the presence of MC in order to emphasize
the effect of the IMF slope.}.
The resulting IRX-$\beta$ diagram is shown in figure
\ref{fig_irxuv_kro}, illustrating that the models well compare
with observed values.

The results show that the IMF slope has an effect on
the final UV colour producing redder models that agree in a better way with 
the observations than set C models.

The larger reddening of these models respect Salpeter IMF based ones is simply
due to the difference in the relative contribution of old (and less obscured) 
and young (and more obscured) stars
to UV luminosity. In fact, young star populations with Kroupa IMF have less
massive stars (and a lower UV luminosity) respect to populations with 
Salpeter IMF; thus, similarly to what happens in the set D models, a smaller
obscuration by the cirrus is needed to make the model redder.

Finally, we computed the attenuation laws of Figure \ref{fig_curv_att} with the Kroupa IMF. 
However we found that there is no visible change in the attenuation laws between models with
the two different IMF (Salpeter and Kroupa) used in this work.


\section{Summary and discussion}
\label{sec_discus}

In previous sections we investigated the role of various physical and
geometrical factors in the relation between the reddening of the UV {\it GALEX}
colour and the attenuation. Our relatively realistic modelling allows a proper
discussion of the effect of the geometrical distribution of stars and the dust, 
and of the effect of the IMF on  the attenuation properties of normal spiral
galaxies. 

The {\it GALEX} data are in principle very sensitive to the dust properties
because the NUV band is strongly affected by the presence of
the 2175 \AA\ bump in the extinction law; e.g. the extinction in NUV is
actually slightly higher than in the FUV for a MW dust. For this reason, such
data could be used to derive the dust properties.

Under the (unrealistic) assumption that stars suffer an extinction independent
of their age (set A, figure \ref{fig_mod_disc}), it is not possible to produce
a reddening in a star population with a MW dust, regardless the relative
geometry of dust and stars. This result confirms the  findings of
\citet{witt00}. Under this assumption, the IRX-$\beta$ diagram for normal
galaxies drawn by \citet{bell02} and \citet{kong04}, and confirmed by
{\it GALEX} data \citep{buat05,cort06,seib05}, would imply a strong reduction
of the 2175 \AA\ bump in the extinction law of these galaxies (see also \citealt{calz05}).

\citet{burg05b} analysed the observed relation in an empirical way, using an
average attenuation law rather than a radiation transport model. Fitting the
UV, the optical SEDs and the
IR luminosity of the UV selected sample, they found that the mean strength of
the 2175 \AA\ bump in the attenuation law characterizing this sample is around
half of the strength in the MW extinction law. However, as the authors
underline, this approach cannot determine if the small bump strength
is due to a dust characteristic or a geometrical effect.

Using our radiation transport model we have shown that, indeed, the observed
relation can be explained without the need of dust properties different from
the MW diffuse medium. This result can be obtained taking into account the fact
that young stars suffer a higher dust attenuation with respect to older stars
(age-dependent extinction).

We have shown (set B, figure \ref{fig_irxuv_mc}) that in the case with all the
dust in molecular clouds obscuring stars younger than a given age, the observed
{\it GALEX} IRX-$\beta$ diagram can be fairly well reproduced. The important
point of this result is that in this case the FUV$-$NUV colour is completely
independent of the extinction law (because the MC are supposed to be optically
thick) so it can be obtained with every kind of dust. It is worth stressing
that such geometrical effect could completely wash out the effect of optical
properties of the dust, and undermine the idea that the IRX-$\beta$ diagram
could be used to derive the dust properties.

However the above case is extreme and not very realistic. Moreover we found
that the timescale on which young stars must be attenuated to explain redder
galaxies is significantly larger (50 Myr or more) than the typical time over
which young stars get rid of their parental clouds (few Myr). Anyway, the good
reproduction of the observed relation suggests that some form of age-dependent
extinction has to be in place.

We interpreted this result as the effect of the different scale-height of young
and old stars in spiral galaxies. By taking into account that the stars are
born near the galactic plane and therefore suffer a higher dust attenuation
than older stars which are distributed in a broader way, we have shown (set C,
figure \ref{fig_irxuv_sal}) that it is possible to obtain a significant
reddening also in the case of a dust with a prominent 2175 \AA\ bump. 
It is worth noting that the well known different distribution of
young and old stars in discs was taken into account only in few population
synthesis models (e.g. \citealt{tuff04,buat96}) but its effect on
the UV colour was never investigated.

The timescale required by new stars to leave the galactic plane needed to
explain
the observed IRX-$\beta$ relation is around 100 Myr. This is in keeping with
observations in our own galaxy, according to which stars younger than $\sim$
150 Myr have a much smaller scale-height than older ones \citep{robi03}.
Moreover, our results (set D, figure \ref{fig_irxuv_sal_mc}) show that the
combination of molecular clouds obscuration and the age dependent scale-height
of stars can reproduce most of the observed IRX-$\beta$.

In this geometry, as the young stars become more and more obscured at
increasing the dust amount, the NUV flux become dominated by the older stars,
while the FUV is still mostly provided by younger ones. Therefore, the
FUV$-$NUV colour is, in this framework, the result of the age dependent
obscuration of the different populations rather than the reddening properties
of the dust. Moreover, our results show that the lack of
correlation between the birthrate parameter and the distance from the
starburst relation could be ascribed at least in part to the variation of
$t_{\rm thin}$, the timescale required by new stars to leave the galactic plan.

We underline that our results do not rule out that the intrinsic
optical properties of dust in different galaxies can be
substantially different from those of the average MW dust. This is
indeed in our view a likely possibility, in general. However, we
have demonstrated that conclusions on this issue can be seriously
affected by the use of too simple models. In particular, the
effects of dust properties in datasets such as those considered in
this work are probably overwhelmed by the age dependent geometry.

We have also shown that the IMF can have a role in shaping the IRX-$\beta$
diagram. Even if the UV colour of a
simple stellar population does not depend on the IMF because it is dominated by
the most massive stars, the rate at which the UV luminosities decrease with
time is given by the IMF slope at higher masses. For this reason, the UV colour
of the integrated SED over the star formation history depends on the IMF.


\section*{Acknowledgements}
P. Panuzzo acknowledges the warm hospitality and financial support of the LAM
for this work.
This work was partially funded by the European Community by means 
of the Marie Curie contract MRTN-CT-2004-503929 "MAGPOP".


\bsp

\label{lastpage}

\end{document}